\documentclass[sigconf]{acmart}

\pdfoutput=1

\AtBeginDocument{%
  \providecommand\BibTeX{{%
    \normalfont B\kern-0.5em{\scshape i\kern-0.25em b}\kern-0.8em\TeX}}}

\copyrightyear{2019}
\acmYear{2019}
\setcopyright{rightsretained}
\acmConference[MMSys '19]{MMSys '19: ACM Multimedia Systems Conference - Demo Track}{June 18--21, 2019}{Amherst, MA, USA}
\acmBooktitle{MMSys '19: ACM Multimedia Systems Conference - Demo Track, June 18--21, 2019, Amherst, MA, USA}
\acmPrice{15.00}
\acmDOI{10.1145/3304109.3323835}
\acmISBN{978-1-4503-6297-9/19/06}

\begin{document}

\title{HTML5 MSE Playback of MPEG 360 VR Tiled Streaming}
\subtitle{JavaScript implementation of MPEG-OMAF viewport-dependent video profile with HEVC tiles}

\author{Dimitri Podborski, Jangwoo Son, Gurdeep Singh Bhullar, Robert Skupin, Yago Sanchez}
\orcid{0000-0003-4726-6685}
\author{Cornelius Hellge, Thomas Schierl}
\affiliation{
  \institution{Fraunhofer Heinrich Hertz Institute}
  \department{Multimedia Communications Group}
  \streetaddress{Einsteinufer 37}
  \city{Berlin}
  \postcode{10587}
  \country{Germany}
}
\email{name.surname@hhi.fraunhofer.de}

\renewcommand{\shortauthors}{Dimitri Podborski, et al.}

\begin{abstract}
  Virtual Reality (VR) and 360-degree video streaming have gained significant attention in recent years. First standards have been published in order to avoid market fragmentation. For instance, 3GPP released its first VR specification to enable 360-degree video streaming over 5G networks which relies on several technologies specified in ISO/IEC 23090-2, also known as MPEG-OMAF. While some implementations of OMAF-compatible players have already been demonstrated at several trade shows, so far, no web browser-based implementations have been presented. In this demo paper we describe a browser-based JavaScript player implementation of the most advanced media profile of OMAF: \textit{HEVC-based viewport-dependent OMAF video profile}, also known as tile-based streaming, with multi-resolution HEVC tiles. We also describe the applied workarounds for the implementation challenges we encountered with state-of-the-art HTML5 browsers. The presented implementation was tested in the Safari browser with support of HEVC video through the HTML5 Media Source Extensions API. In addition, the WebGL API was used for rendering, using region-wise packing metadata as defined in OMAF.
\end{abstract}



\maketitle

\section{Introduction}
Virtual Reality (VR) and 360-degree video streaming have gained popularity among researchers and the multimedia industry in recent years. For example, in addition to many published research papers in this area, several standardization organizations such as ISO/IEC, MPEG and 3GPP have published their first specifications on VR \cite{3gpp} \cite{omaf}. One such specification is the result of an MPEG activity and is referred to as the Omnidirectional Media Format (OMAF) that specifies a storage and delivery format for 360-degree multimedia content. OMAF particularly defines the HEVC-based viewport-dependent media profile for video which allows to stream HEVC tiles of different resolutions and finally combine them into a single bitstream so that only one video is decoded on the client end-device. This approach, often referred to as tile-based streaming, allows the service operator to increase the resolution within the viewport while decoding a lower resolution video compared to the conventional naive 360-degree video streaming approach. However, this is coupled with additional complexity of the player implementation, particularly when implemented in a web browser environment using only JavaScript. In addition, the W3C Media Source Extensions (MSE) offer no direct support for OMAF which requires workarounds at certain functional steps. In this paper we describe how to overcome these challenges in order to implement a fully standard-compliant OMAF player for tile-based streaming in JavaScript. Furthermore, the entire source code of our implementation is available on GitHub \cite{omafjs}.

The remainder of the paper is structured as follows. Section 2 presents the overall architecture of our implementation and describes each component of the system. The proof of concept is presented in Section 3, explaining the most important challenges and workarounds. Finally, Section 4 concludes our paper.

\section{Architecture}
A general overview of the main components involved in our implementation is depicted in Figure 1. It consists of six main modules which interact with each other and together provide a fully functional player for OMAF  360-degree video. Those modules are: \textbf{Player}, Downloader (\textbf{DL}), MPD Parser (\textbf{MP}), Scheduler (\textbf{SE}), Media Engine (\textbf{ME}) and finally the Renderer (\textbf{RE}).
\begin{figure}[h]
  \centering
  \includegraphics[width=\linewidth]{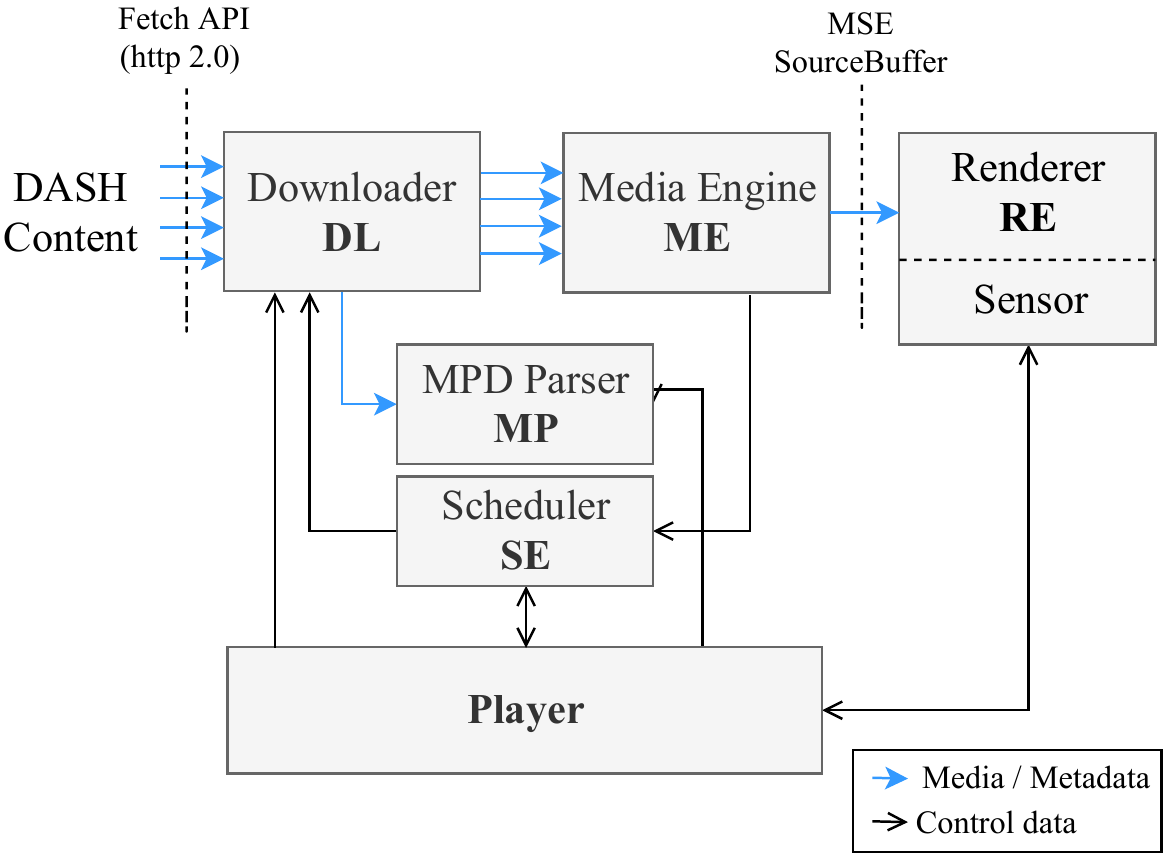}
  \caption{Overall architecture.}
  \Description{Overall architecture.}
\end{figure}

The \textbf{Player} module represents the core of the entire application. It connects all modules with each other and controls them. The \textbf{DL} module deals with all HTTP requests to the server. The \textbf{MP} module implements the parsing of the DASH manifest file (MPD) together with additional metadata defined in OMAF. The \textbf{SE} module controls the DL module and decides when requests for the next segments should be executed, based on the current status of the player. The task of the \textbf{ME} module is to parse OMAF related metadata on the File Format level and re-package the downloaded OMAF content in such a way that the Media Source Extensions API of the web browser can process the data. Finally, the \textbf{RE} module uses the OMAF metadata in order to correctly render the video texture on the canvas using WebGL API. The following subsections describe each of these six modules in more detail.

\subsection{Player}
As already mentioned in the previous section the Player module can be seen as the core of the entire application. Its main goal is to connect all modules with each other and to control them. It also connects HTML elements such as video and canvas from the main page with the ME and RE modules. In addition, it provides a basic functionality to the user such as load a source, play, pause, loop, change into or go out of full screen mode and retrieve certain metrics of the application in order to plot the data on the screen.

\subsection{Downloader}
The main task of this module is to manage all HTTP traffic between our application and the server. This module receives a list of URLs from the player module and downloads them using the Fetch API. After all required HTTP requests are processed and all requested data is successfully downloaded, it forwards the data to the ME module for further processing. Since the current version of the player fires a lot of simultaneous HTTP requests it is desirable to host the media data on an HTTP/2 enabled server in order to improve the streaming performance. 

\subsection{MPD Parser}
OMAF uses Dynamic Adaptive Streaming over HTTP (DASH) as a primary delivery mechanism for VR media. It also specifies additional metadata for 360-degree video streaming such as:
\begin{itemize}
  \item \textit{Projection type}: only Equirectangular (ERP) or Cubemap (CMP) projections are allowed.
  \item \textit{Content coverage}: is used to determine which region each DASH Adaptation Set covers while each HEVC tile is stored in a separate Adaptation Set. This information is required to select only those Adaptation Sets which cover the entire 360-degree space.
  \item \textit{Spherical region-wise quality ranking}: (SRQR) is used to determine where the region with the highest quality is located within an Adaptation Set. This metadata allows us to select an Adaptation Set based on current orientation of the viewport.
\end{itemize}
In addition to OMAF metadata, another notable feature is the DASH Preselection Descriptor which indicates the dependencies between different DASH Adaptation Sets.
The MP module parses all required DASH manifest metadata and implements several helper functions which are used by the Player module in order to make appropriate HTTP requests.

\subsection{Scheduler}
One of the key aspects of any streaming service is maintaining a sufficient buffer in order to facilitate smooth media playback. In the implementation, the buffer is maintained using a parameter named 'buffer limit' which can be set prior or during a streaming session. The parameter indicates the maximum buffer fullness level in milliseconds and depending on its value the SE module schedules the next request. If the buffer is able to accommodate a segment, then the SE module initiates the request for the next segment. On the other hand, if the buffer is full, then the request for the next segment is delayed until buffer fullness and the current time of the media playback indicate otherwise. 

An important point to mention for any viewport-dependent streaming implementation, is that the user can change the viewport orientation during the playback at any time. Therefore, the system should adapt to the changed viewport orientation as quickly as possible, which implies that the buffer fullness limit must be kept relatively small. Preferably, it should be in the range of a few seconds.

\subsection{Media Engine}
The ME module is taking input from the DL module and prepares the downloaded media segments for the consumption by the Media Source Extension (MSE). Current state-of-the-art MSE implementations do not yet fully support all modern features of the ISO Base Media File Format, in particular some File Format features used in OMAF media segments. Therefore, the downloaded media segment data is repackaged in a format that is compatible with MSE implementations. For the parsing and writing of the File Format data we are using the JavaScript version of GPAC's MP4Box tool mp4box.js \cite{gpac}. Furthermore, the parsing of the OMAF related metadata is also implemented in the ME module. Figure 2 visualizes the repackaging process and shows the flow of downloaded media segments through the ME module to the MSE SourceBuffer.
\begin{figure}[h]
  \centering
  \includegraphics[width=\linewidth]{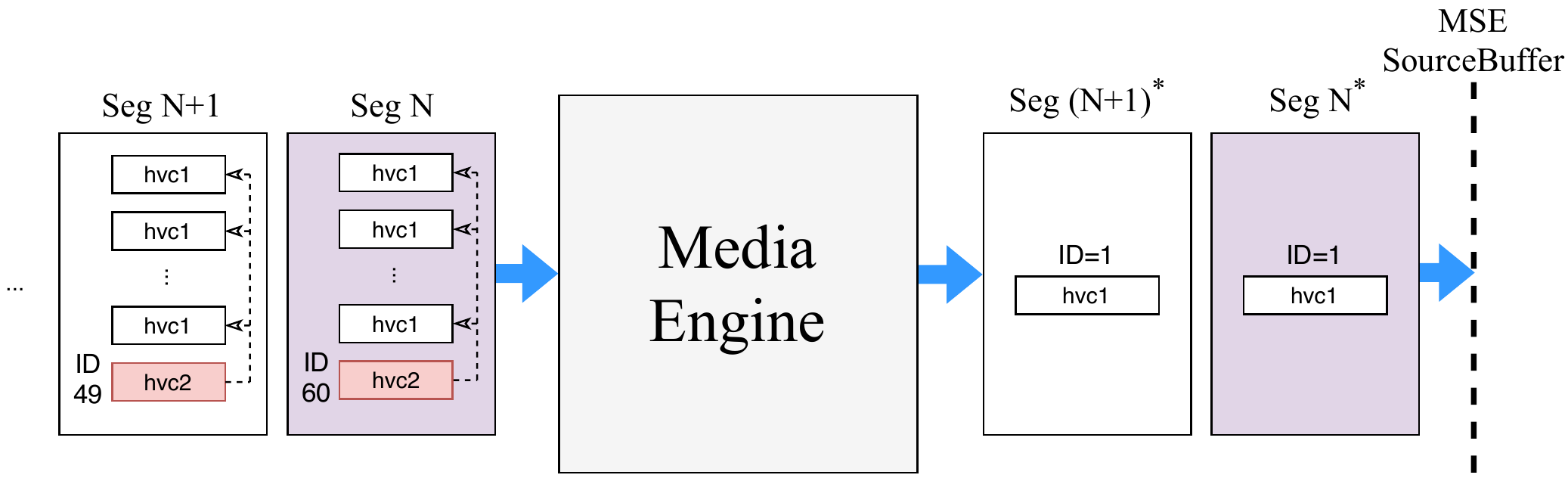}
  \caption{Media Engine module repackages the downloaded DASH segments before pushing them to MSE SourceBuffer.}
  \Description{Media Engine repackaging process.}
\end{figure}

Before the ME module processes the media segments, each segment should be completely downloaded and consist of multiple hvc1 video tracks (one track for each HEVC tile) and one additional hvc2 video track with Extractor NAL Units for HEVC video \cite{nal}. The extractor track is required for the creation of a single HEVC bitstream from the individually delivered HEVC tiles in order to use a single decoder instance. An extractor represents the in-stream data structure using a NAL unit header for extraction of data from other tracks and can be logically seen as a pointer to data located in a different File Format track. Unfortunately, currently available web browsers do not natively support File Format extractor tracks and thus a repackaging workaround as performed by the ME module is necessary. Therefore, the ME module resolves all extractors within an extractor track and packages the resolved bitstream into a new track with a unique track ID, even if the extractor track ID changes. Hence, from the MSE SourceBuffer perspective, it looks like the segments are coming from the same DASH Adaptation Set even if the player switches between different tiling configurations.

\subsection{Renderer}
After the repackaged segment is processed by the MSE SourceBuffer, the browser decodes the video and the video texture is finally rendered by the RE module using OMAF metadata.  The RE module is implemented using a custom shader written in OpenGL Shading Language (GLSL) together with a three.js library (WebGL library) \cite{threejs} which is used to implement three-dimensional graphics on the Web. 
Our rendering implementation is based on triangular polygon mesh objects and supports both:  equirectangular and cubemap pojections. In case of a cubemap projection one cube face is divided into two triangular surfaces, while in case of an equirectangular projection, a helper class from three.js library is used to create a sphere geometry.

Figure 3 shows three main processes used for rendering, from the decoded picture to the final result rendered on the cube. It shows an example where each face of the cube is divided into four tiles while the decoded picture is composed of 12 high-resolution and 12 low-resolution tiles. The 12 triangular surfaces of the cube as depicted in Figure 3 (c) can be represented as a 2D plane like in Figure 3 (b). The fragment shader of the RE module uses OMAF metadata to render the decoded picture correctly at the cube faces as shown in Figure 3 (b). The Region-wise Packing (RWPK) of OMAF metadata has top-left position, width, height in packed and unpacked coordinates as well as rotation of tiles for all tracks.
\begin{figure}[h]
  \centering
  \includegraphics[width=\linewidth]{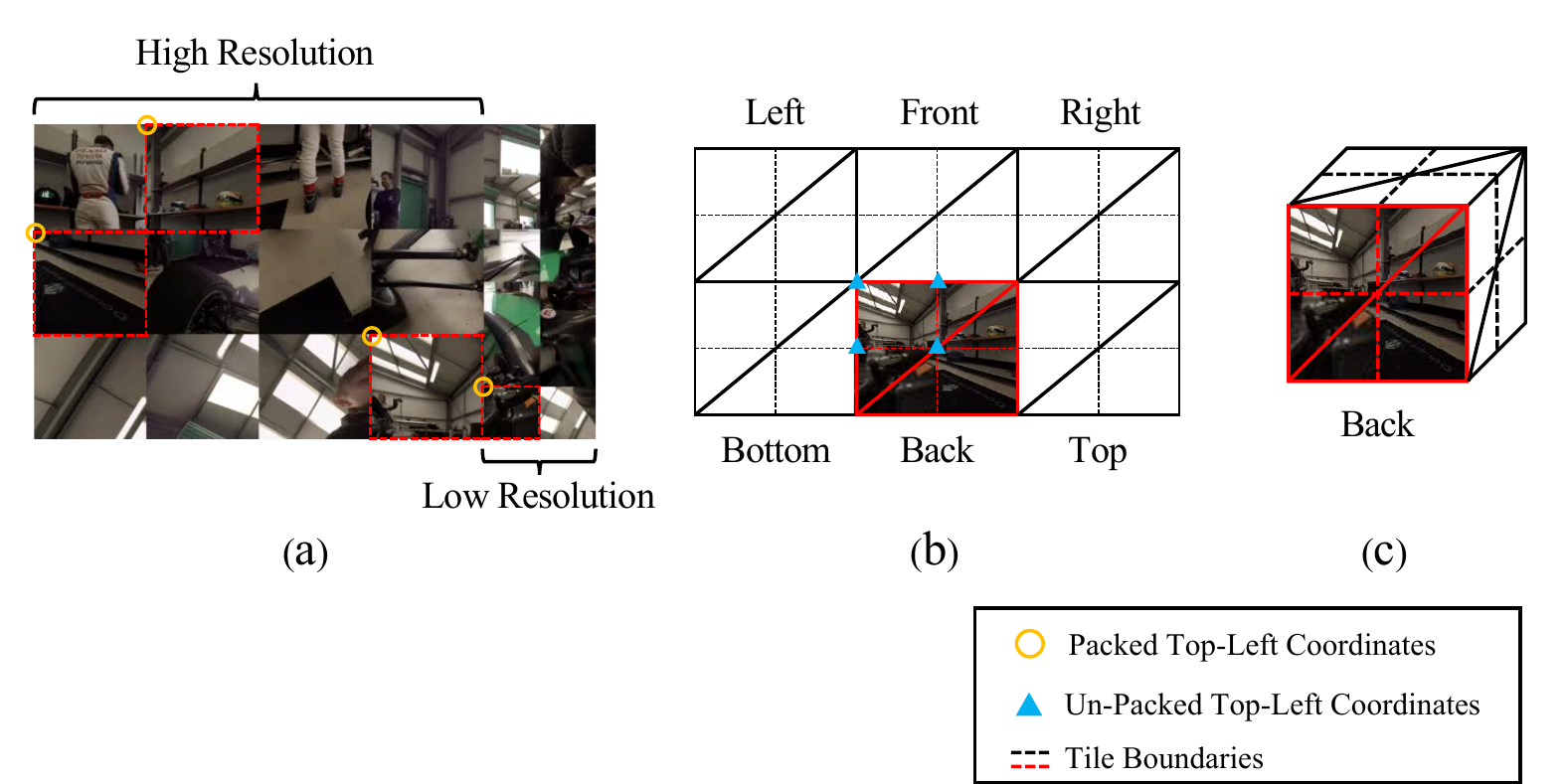}
  \caption{Rendering process of one face of the cube; (a) the decoded video texture before unpacking; (b) the 12 triangular surfaces on the 2D plane after region-wise unpacking; (c) the composition of 12 triangular surfaces on the cube.}
  \Description{Rendering process.}
\end{figure}
Since the shader reconstructs the position of pixels based on OMAF metadata, Figure 3 (b) can be assumed to be a region-wise Un-Packed image. Therefore, the shader sets the rendering range of the Figure 3 (a) using the RWPK metadata, and renders the tiles of the decoded picture to the cube faces of the Figure 3 (b). However, when there is a change in the viewport position, the RE module has to be given correct metadata for the current track. In the implementation, when the manifest file is loaded, the RE module is initialized with all RWPK metadata to correctly render all tracks. The synchronization of the track switching is covered in the following section.

\section{Proof of concept}
In this section, we first give a brief overview of the implementation. We then discuss the most important challenges we encountered during implementation and describe their workarounds.

\subsection{Implementation overview}
Figure 4 shows the screenshot of the main page of our OMAF JavaScript Player. In addition to some basic video player controls such as load MPD, play, pause, change to full screen mode etc., it also provides several controls for debugging purposes.
In the top input menu, the user can provide the URL to a DASH manifest file (MPD) and load it.
\begin{figure}[h]
  \centering
  \includegraphics[width=\linewidth]{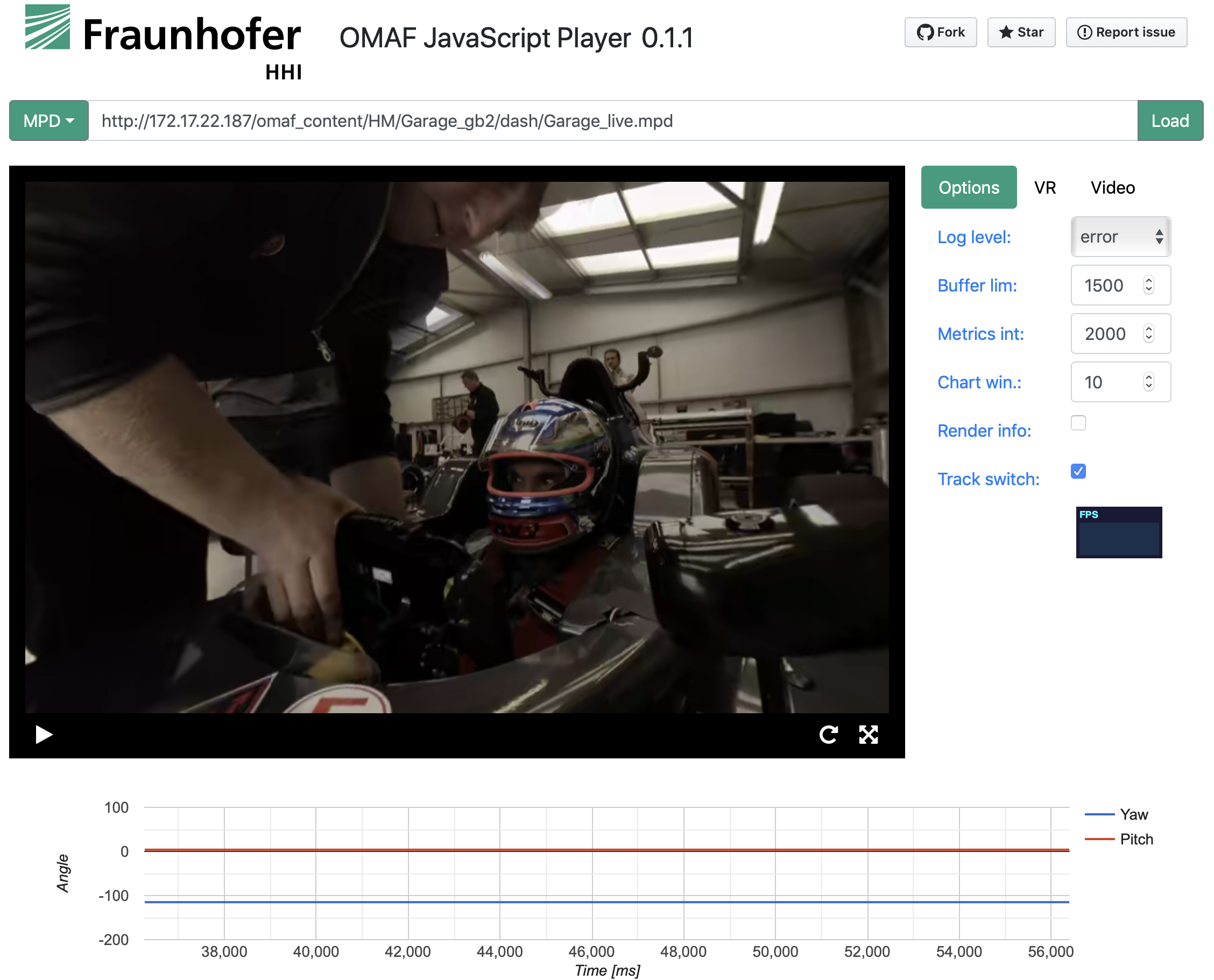}
  \caption{Screenshot of the OMAF JavaScript Player user interface.}
  \Description{OMAF JavaScript Player user interface.}
\end{figure}
After the MPD is successfully loaded and parsed, the player downloads all initialization segments (one for each emphasized viewport, which, in relation to the example in Figure 3, corresponds to 24 extractor tracks), parses OMAF-related metadata and initializes the RE module with extracted region-wise packing information.
In addition, the ME module creates a pseudo-initialization segment for the MSE SourceBuffer to initialize it in a way such that following repackaged media segments can be successfully decoded by the web browser. 

The streaming session starts when the user presses the play button. The player then continuously downloads media segments of a certain extractor track, depending on the current orientation of the viewport. In addition, all dependent media segments (hvc1 tracks) are also downloaded. All downloaded media segments are then immediately repackaged and the corresponding RWPK metadata is used to correctly render the video on the canvas. We tested our implementation on Safari 12.02 since only web browsers from Apple and Microsoft\footnote{Microsoft Edge web browser supports the implementation on devices with a hardware HEVC decoder. For devices that do not provide hardware support for HEVC, the HEVC Video Extensions have to be enabed in the Microsoft Store.} natively support it. Raw video sequences were provided by Ericsson and prepared for streaming using the OMAF file creation tools from Fraunhofer HHI \cite{hhitools} \cite{omafjs}. Finally, the content was hosted on Amazon CloudFront CDN, which supports HTTP/2 in conjunction with HTTPS.

The following section covers some of the issues that we faced during the implementation.

\subsection{Implementation challenges}
An important prerequisite for good functionality of the implementation is the synchronization of the extractor track switching and the corresponding RWPK metadata. When the user changes the viewport, the extractor track is changed and the high and low-quality tile positions of the decoded picture are derived using the corresponding region-wise packing metadata. The RE module has to reset the texture mapping of the decoded picture according to the changed track and it shall be done at exactly the same time when the video texture changes from one track to another. The simplest way to detect the exact frame number is to check the current time of the video. Unfortunately, W3C organizations are still discussing the precise accuracy of the currentTime of the video element \cite{w3c} and nowadays it is not possible to conveniently detect the exact frame number of the video reliably using currentTime. Therefore, the ME module uses two video elements together with two MSE SourceBuffers and alternately switches between them when the extractor track (and RWPK) changes. The ME module saves the bitstream of the changed track in the SourceBuffer of the other video element. When the Player reaches the end of the active buffer it subsequently switches to the other video element. At the same time, the player informs the RE module through an event about the change of the video element. The RE module declares two scene objects and associates them with each video element. Also, the RE module calculates and stores the RWPK metadata of the decoded picture for all tracks in the initialization phase. When receiving the event about the change of the video element from the Player, the RE module replaces the scene and maps the video texture of the decoded picture based on the scene object, so that track synchronization is performed without an error.

While this solution works well on Safari, we discovered an open issue on Microsoft Edge browser \cite{edge} that interferes with the two buffers workaround. The Edge web browser requires a few seconds of data in each buffer in order to start the decoding process, and therefore the new segment at every track switch cannot be instantly rendered.

Furthermore, due to the track synchronization solution using two video elements, we need to operate two MSE SourceBuffer objects, which makes the buffering logic a bit more complex as the SE module has to monitor the buffer fullness level of both SourceBuffer objects. The duration of media segments present in the two media sources is combined together to determine the available buffer time at a given moment so that the SE module can make requests for the future media segments accordingly.

For future work we plan to further optimize the streaming performance of the player while reducing the amount of HTTP requests and implement suitable rate-adaptation algorithms for tile-based streaming.

\section{Conclusions}
In this paper, we present the first implementation for the playback of the \textit{HEVC-based viewport-dependent OMAF video profile} in a web browser. 
This OMAF profile allows to increase the resolution of 360-degree video inside the viewport by sacrificing resolution of the areas which are not presented to the user. The entire implementation is in JavaScript and therefore no additional plugins are required. Although current state-of-the-art MSE implementations do not yet fully support some File Format features used in OMAF, our implementation shows how these limitations can be overcome by using JavaScript in combination with MSE. We also point out that W3C is currently working on accurate frame-by-frame signaling of media elements, which currently requires a slightly more complex workaround in the rendering process using two MSE SourceBuffer elements in alternating order. The entire source code of the implementation presented in this paper is published on GitHub \cite{omafjs}. Information on other MPEG-OMAF implementations from Fraunhofer HHI can be found in \cite{hhitools}.

\bibliographystyle{ACM-Reference-Format}
\bibliography{ms}


\begin{thebibliography}{9}


\ifx \showCODEN    \undefined \def \showCODEN     #1{\unskip}     \fi
\ifx \showDOI      \undefined \def \showDOI       #1{#1}\fi
\ifx \showISBNx    \undefined \def \showISBNx     #1{\unskip}     \fi
\ifx \showISBNxiii \undefined \def \showISBNxiii  #1{\unskip}     \fi
\ifx \showISSN     \undefined \def \showISSN      #1{\unskip}     \fi
\ifx \showLCCN     \undefined \def \showLCCN      #1{\unskip}     \fi
\ifx \shownote     \undefined \def \shownote      #1{#1}          \fi
\ifx \showarticletitle \undefined \def \showarticletitle #1{#1}   \fi
\ifx \showURL      \undefined \def \showURL       {\relax}        \fi
\providecommand\bibfield[2]{#2}
\providecommand\bibinfo[2]{#2}
\providecommand\natexlab[1]{#1}
\providecommand\showeprint[2][]{arXiv:#2}

\bibitem[\protect\citeauthoryear{3GPP}{3GPP}{2019}]%
        {3gpp}
\bibfield{author}{\bibinfo{person}{3GPP}.} \bibinfo{year}{2019}\natexlab{}.
\newblock \bibinfo{booktitle}{\emph{{5G; 3GPP Virtual reality profiles for
  streaming applications}}}.
\newblock \bibinfo{type}{Technical Specification (TS)} 26.118.
  \bibinfo{institution}{{3rd Generation Partnership Project}}.
\newblock
\newblock
\shownote{Version 15.1.0.}


\bibitem[\protect\citeauthoryear{GPAC}{GPAC}{2019}]%
        {gpac}
GPAC \bibinfo{year}{2019}\natexlab{}.
\newblock \bibinfo{title}{GPAC Multimedia Open Source Project, JavaScript
  version of GPACs MP4Box tool}.
\newblock
\newblock
\urldef\tempurl%
\url{https://gpac.github.io/mp4box.js}
\showURL{%
Retrieved April 19, 2019 from \tempurl}


\bibitem[\protect\citeauthoryear{HHI}{HHI}{2019a}]%
        {hhitools}
\bibfield{author}{\bibinfo{person}{Fraunhofer HHI}.}
  \bibinfo{year}{2019}\natexlab{a}.
\newblock \bibinfo{title}{Better quality for 360-degree video}.
\newblock
\newblock
\urldef\tempurl%
\url{http://hhi.fraunhofer.de/omaf}
\showURL{%
Retrieved April 19, 2019 from \tempurl}


\bibitem[\protect\citeauthoryear{HHI}{HHI}{2019b}]%
        {omafjs}
\bibfield{author}{\bibinfo{person}{Fraunhofer HHI}.}
  \bibinfo{year}{2019}\natexlab{b}.
\newblock \bibinfo{title}{HTML5 MSE Playback of MPEG 360 VR Tiled Streaming}.
\newblock
\newblock
\urldef\tempurl%
\url{https://github.com/fraunhoferhhi/omaf.js}
\showURL{%
Retrieved April 19, 2019 from \tempurl}


\bibitem[\protect\citeauthoryear{ISO/IEC}{ISO/IEC}{2017}]%
        {nal}
ISO/IEC \bibinfo{year}{2017}\natexlab{}.
\newblock \bibinfo{booktitle}{\emph{14496-15, Information technology - Coding
  of audio-visual objects - Part 15: Carriage of network abstraction layer
  (NAL) unit structured video in the ISO base media file format}}.
\newblock ISO/IEC.
\newblock


\bibitem[\protect\citeauthoryear{ISO/IEC}{ISO/IEC}{2019}]%
        {omaf}
ISO/IEC \bibinfo{year}{2019}\natexlab{}.
\newblock \bibinfo{booktitle}{\emph{23090-2, Information technology - coded
  representation of immersive media (MPEG-I) - Part 2: Omnidirectional media
  format.}}
\newblock ISO/IEC.
\newblock


\bibitem[\protect\citeauthoryear{Media and Group}{Media and Group}{2018}]%
        {w3c}
\bibfield{author}{\bibinfo{person}{W3C Media} {and}
  \bibinfo{person}{Entertainment~Interest Group}.}
  \bibinfo{year}{2018}\natexlab{}.
\newblock \bibinfo{title}{Frame accurate seeking of HTML5 MediaElement}.
\newblock
\newblock
\urldef\tempurl%
\url{https://github.com/w3c/media-and-entertainment/issues/4}
\showURL{%
Retrieved April 19, 2019 from \tempurl}


\bibitem[\protect\citeauthoryear{three.js}{three.js}{2019}]%
        {threejs}
three.js \bibinfo{year}{2019}\natexlab{}.
\newblock \bibinfo{title}{Three.js, JavaScript 3D Library}.
\newblock
\newblock
\urldef\tempurl%
\url{https://threejs.org/}
\showURL{%
Retrieved April 19, 2019 from \tempurl}
\newblock
\shownote{Version r101.}


\bibitem[\protect\citeauthoryear{V.}{V.}{2017}]%
        {edge}
\bibfield{author}{\bibinfo{person}{David V.}} \bibinfo{year}{2017}\natexlab{}.
\newblock \bibinfo{title}{Video MSE issues as of March 01 2017}.
\newblock
\newblock
\urldef\tempurl%
\url{https://developer.microsoft.com/en-us/microsoft-edge/platform/issues/11147314}
\showURL{%
Retrieved April 19, 2019 from \tempurl}


\end{thebibliography}

\end{document}